\documentclass[pra,reprint,superscriptaddress,longbibliography]{revtex4-2}

\usepackage{array}
\usepackage{natbib}
\usepackage[section]{placeins}
\usepackage{pgfplots,amsfonts}
\usepackage[breaklinks]{hyperref}
\usepackage{amssymb}
\usepackage{amsmath,amsthm,bm}
\usepackage{amsfonts}
\usepackage{cleveref}
\usepackage{verbatim}
\usepackage{mathtools}
\usepackage{tikz}
\hypersetup{citecolor=red,colorlinks=true,urlcolor=blue}
\usepackage{algorithm,setspace}
\usepackage{algpseudocode}
\usetikzlibrary{angles, quotes}
\usetikzlibrary{positioning}

\usepackage[utf8]{inputenc} 
\DeclareUnicodeCharacter{0308}{\"{}} 
\DeclareUnicodeCharacter{03C0}{$\pi$} 

\usepackage[normalem]{ulem}
\usepackage{color}
\usepackage{xcolor}

\newcolumntype{P}[1]{>{\centering\arraybackslash}p{#1}}

\newcommand{\Tr}{\mbox{Tr}}

\newcommand{\bra}[1]{\mbox{$\langle #1 |$}}
\newcommand{\ket}[1]{\mbox{$| #1 \rangle$}}

\definecolor{Ugreen}{HTML}{198a11}

\begin{document}
\title{Quantum Many-body Simulations from a Reinforcement-Learned Exponential Ansatz}

\author{Yuchen Wang and David A. Mazziotti}

\email{damazz@uchicago.edu}

\affiliation{Department of Chemistry and The James Franck Institute, The University of Chicago, Chicago, Illinois 60637, USA}

\date{Submitted May 3, 2025}

\begin{abstract}
Solving for the many-body wavefunction represents a significant challenge on both classical and quantum devices because of the exponential scaling of the Hilbert space with system size.  While the complexity of the wavefunction can be reduced through conventional ans\"{a}tze (e.g., the coupled cluster ansatz), it can still grow rapidly with system size even on quantum devices.  An exact, universal two-body exponential ansatz for the many-body wavefunction has been shown to be generated from the solution of the contracted Schr{\"o}dinger equation (CSE), and recently, this ansatz has been implemented without classical approximation on quantum simulators and devices for the scalable simulation of many-body quantum systems.  Here we combine the solution of the CSE with a form of artificial intelligence known as reinforcement learning (RL) to generate highly compact circuits that implement this ansatz without sacrificing accuracy.  As a natural extension of CSE, we reformulate the wavefunction update as a Markovian decision process and train the agent to select the optimal actions at each iteration based upon only the current CSE residual.  Compact circuits with high accuracy are achieved for H\(_3\) and H\(_4\) molecules over a range of molecular geometries.
\end{abstract}

\maketitle

\section{Introduction}

Quantum computers have emerged as promising platforms for solving complex quantum chemistry problems that can become intractable for classical computers~\cite{lanyon2010, o2016, mcardle2020, dutta2024}. Hybrid quantum-classical algorithms, such as the variational quantum eigensolver (VQE)~\cite{cerezo2021, kandala2017, tilly2022}, have demonstrated notable applications in chemistry using near-term quantum devices. The VQE optimizes a parameterized quantum circuit to approximate the system energy by leveraging quantum hardware for state preparation and measurement while using classical computers for parameter optimization. A challenge for VQEs lies in designing ansatz circuits that are both highly expressive to capture electron correlation accurately and sufficiently shallow to execute reliably on noisy quantum hardware, while also being capable of navigating complex optimization landscapes~\cite{wecker2015,d2023,larocca2024}.


Given the substantial challenges of VQE, we have developed an alternative approach to solving the Schr\"{o}dinger equation on quantum computers that does not rely upon a fixed parametrization, known as the contracted quantum eigensolver (CQE)~\cite{smart2021, smart2022, Boyn.2021u94, Smart.2022w8u, wang2023boson, Wang.20232b, benavides2024, Smart.2024, Smart.2024w8, warren2024, warren2025, wang2025}. CQE iteratively updates the wavefunction based on measuring the residual of the two-particle contracted Schr{\"o}dinger equation (CSE)~\cite{mazziottiContractedSchrodingerEquation1998, Colmenero.1993, Nakatsuji.1996, Mazziotti.1999j9j, Mukherjee.2001, Yasuda.2002, Mazziotti.2002, cohenHierarchyEquationsReduced1976, Valdemoro2007, Mazziotti.20060v3, Mazziotti.2007, Mazziotti.2007k2h, Boyn.2021}---a projection of the Schr{\"o}dinger equation onto the two-electron space---an example of a protocol for partial quantum state tomography~\cite{bonet2020}.  The solution of the CSE has been shown to generate an exact, universal two-body exponential ansatz for the many-body wavefunction~\cite{mazziotti2004exactness, mazziotti2020} with applications on classical computers~\cite{Gidofalvi.2009, Rothman.2009, Snyder.2011u3, Sand.2015, Boyn.2021} as well as quantum simulators and devices~\cite{smart2022, Boyn.2021u94, warren2024, warren2025, wang2025}.  Significantly, the CQE can be viewed as a sequential decision making process, in which an agent selects actions to maximize a defined reward based on feedback from real-time measurement outcomes. Such tasks are inherently well-suited for a specific type of machine learning known as reinforcement learning (RL)~\cite{Kaelbling.1996, Hasselt.2016, Silver.2018, Vinyals.2019}  RL is recognized as a powerful tool in the field of quantum information science with applications in diverse areas such as control optimization~\cite{niu2019,liang2024}, circuit compilation~\cite{NEURIPS2021_97244127, zhang2020}, error correction~\cite{nautrup2019, andreasson2019, olle2024}, and quantum metrology~\cite{fallani2022, belliardo2024}.

In this work we focus on using RL to optimize the many-electron wavefunction ansatz for quantum simulations in molecular science. By treating the CQE ansatz design as a Markovian decision process, we can define agents can learn to minimize circuit depth without sacrificing the exactness of the CQE ansatz for solving many-particle quantum systems.  While this RL method is directly inspired by CQE, it provides guidance for the general ansatz construction problem with applicability to other classes of quantum eigensolvers and algorithms including VQE.  This paper is organized as follows: we first briefly recap the CQE algorithm and establish its essential connection to RL. We then present results based on a deep Q-network implementation of RL for the CQE. Lastly, we provide some discussion and outlook focusing on model transferability.

\section{Theory}

\subsection{Reduced density matrices and the contracted quantum eigensolver}

The $p$-particle reduced density matrix ($p$-RDM) provide a compact representation of the solution to the quantum many-body problem with $p$-body interactions. For molecular Hamiltonians involving at most two-body interactions, the energy is a linear function of the 2-RDM $^2D$~\cite{Coleman.1963}
\begin{equation}
    E = \Tr[^2K ^2D],
\end{equation}
where
\begin{align}
^{2}{K}^{ij}_{kl} & = { \frac{1}{N-1} h^{j}_{l} \wedge \delta^{i}_{k} + \frac{1}{2} u^{ij}_{kl} },\\
^2D^{ij}_{kl} & = \bra{\Psi}\hat{a}^{\dagger}_i\hat{a}^{\dagger}_j\hat{a}^{}_l\hat{a}^{}_k\ket{\Psi} .
\end{align}
Here $\hat{a}_i^{\dagger}$ and $\hat{a}_i$ are the creation and annihilation operators for the $i^{\rm th}$ orbital, $h^{j}_{l}$ and $u^{ij}_{kl}$ are the one- and two-electron integrals, respectively, and $\wedge$ is the Grassmann wedge product~\cite{mazziottiContractedSchrodingerEquation1998, Yokonuma.1992}. For convenience, we write the reduced density operator as $\hat{\Gamma}^{ij}_{kl} = \hat{a}^{\dagger}_i\hat{a}^{\dagger}_j\hat{a}^{}_l\hat{a}^{}_k$ for the remainder of the paper.

Using the 2-RDM as a fundamental variable in solving the Schr\"{o}dinger equation can be advantageous on both classical and quantum devices due to its polynomial $r^{4}$ scaling relative to exponential scaling $\exp{(r)}$ of the wavefunction~\cite{mazziottiQuantumChemistryWave2006, ch8}.  To obtain the fundamental stationary-state equation of the 2-RDM, we contract the Schr\"{o}dinger equation onto the space of two electrons, resulting in the CSE~\cite{mazziottiContractedSchrodingerEquation1998, Colmenero.1993, Nakatsuji.1996, Mazziotti.1999j9j, Mukherjee.2001, Yasuda.2002, Mazziotti.2002, cohenHierarchyEquationsReduced1976, Valdemoro2007},
\begin{equation}\label{eq:cse}
\bra{\Psi}\hat{\Gamma}^{ij}_{kl}(\hat{H}-E)\ket{\Psi} = 0.
\end{equation}
A wavefunction satisfies the CSE if and only if it satisfies the Schr\"{o}dinger equation~\cite{nakatsujiEquationDirectDetermination1976, mazziottiContractedSchrodingerEquation1998}.  The proof of this result in second quantization~\cite{mazziottiContractedSchrodingerEquation1998} follows from showing that the satisfaction of the CSE implies the vanishing of the energy variance which implies the satisfaction of the Schr\"{o}dinger equation.

\begin{figure}[htbp]
    \centering
    \includegraphics[width=0.9\linewidth]{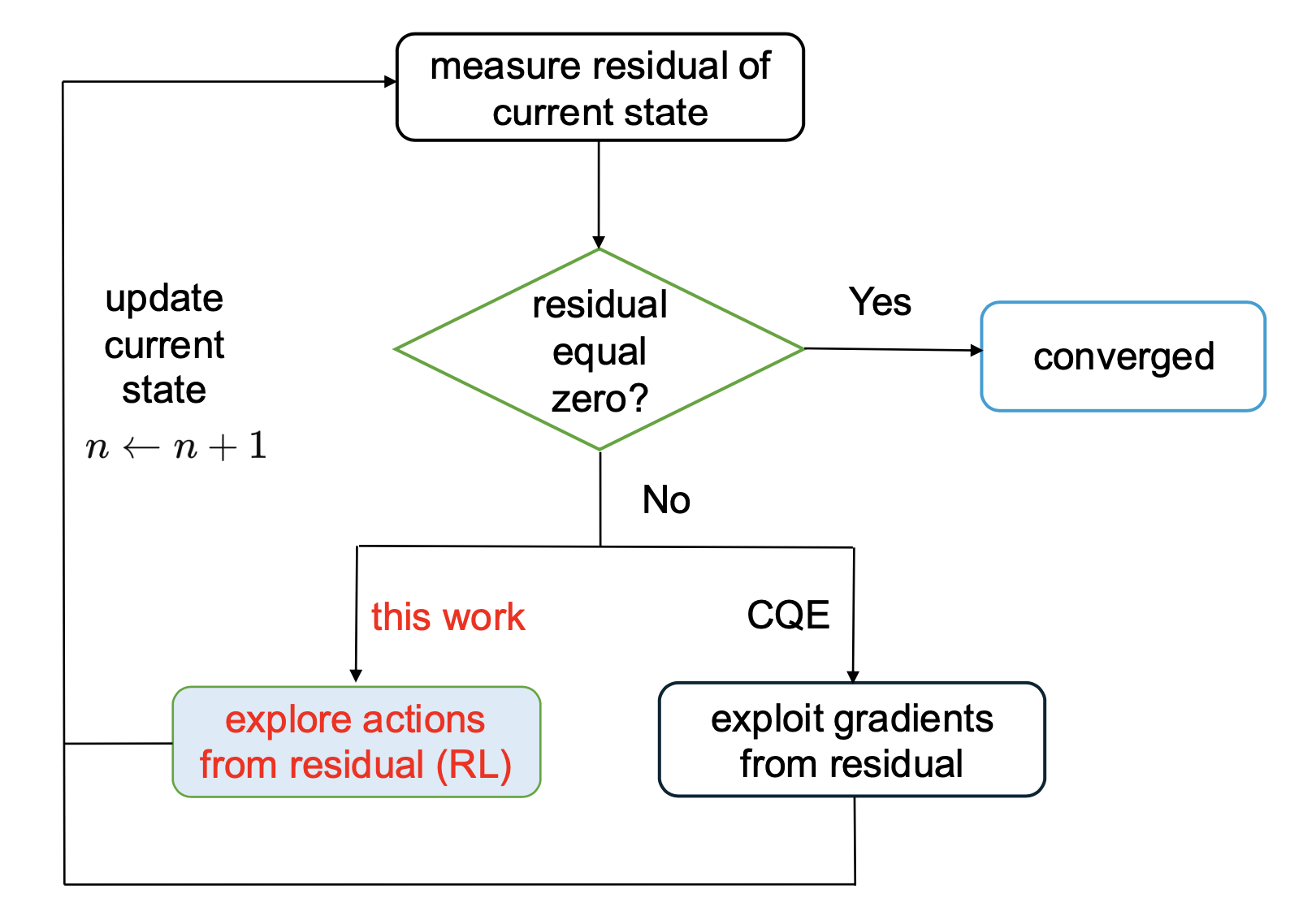}
    \caption{A schematic overview of original CQE algorithm and RL protocol reported in this work.}
    \label{fig:RL}
\end{figure}

While most algorithms for many-electron simulations on quantum devices aim to solve the Schr{\"o}dinger equation, we have developed a family of contracted quantum eigensolvers~\cite{smart2021, smart2022, Boyn.2021u94, Smart.2022w8u, wang2023boson, Wang.20232b, benavides2024, Smart.2024, Smart.2024w8, warren2024, warren2025, wang2025} that solve the CSE.  Solution of the CSE has an advantage relative to the Schr{\"o}dinger equation because its two-body contraction reveals an exact ansatz for the many-particle wavefunction as a product of two-body exponential transformations~\cite{mazziotti2004exactness, mazziotti2020, warren2024}.  Given an arbitrary wavefunction $\ket{\Psi}$, we define the residual of the CSE
\begin{equation}
    \hat{R} = \sum_{ijkl}{}^2\!R^{ij}_{kl}\hat{\Gamma}^{ij}_{kl},
    \label{eq:residual1}
\end{equation}
where ${}^2\!R^{ij}_{kl}$ is decomposable into Hermitian and anti-Hermitian parts
\begin{multline}
    ^2 R^{ij}_{kl} = \bra{\Psi}\hat{\Gamma}^{ij}_{kl}(\hat{H}-E)\ket{\Psi} \\ =\frac{1}{2}( \bra{\Psi}\{\hat{\Gamma}^{ij}_{kl}, \hat{H}-E\}\ket{\Psi} + \bra{\Psi}[\hat{\Gamma}^{ij}_{kl},\hat{H}]\ket{\Psi})
    \label{eq:residual}
\end{multline}
The iterative update of the wavefunction by an exponential transformation defined with the residual operator, as shown in Fig.~\ref{fig:RL}, converges to the exact ground-state wavefunction~\cite{Smart.2024, warren2024}.  Similarly, the excited-state energy and wavefunction can be obtained by minimizing the energy variance~\cite{Wang.20232b}.  A significant benefit of using quantum computers to solve the CSE is that the full residual $\hat{R}$ can be directly measured through partial quantum state tomography. Iterative application of the transformation $\exp{(\theta \hat{R})}$ on the quantum computer converges the wavefunction $\ket{\Psi}$ with $\theta$ serving as a learning rate that can be either fixed~\cite{benavides2024} or dynamically adjusted~\cite{smart2021}. Moreover, the CSE residual can be restricted to its anti-Hermitian portion, known as the ACSE~\cite{Mazziotti.20060v3,Mazziotti.2007, Mazziotti.2007k2h, Boyn.2021}, allowing the transformations to be unitary without dilation~\cite{smart2021, Smart.2022w8u, wang2023boson, benavides2024}.


\subsection{Reinforcement learning approach}

CQE offers an on-the-fly approach to generating a wavefunction ansatz for many-body quantum systems, providing a natural framework for the incorporation of RL techniques~\cite{Kaelbling.1996, Hasselt.2016, Silver.2018, Vinyals.2019}.  First, because the CQE's pool of operators---the set of two-body exponential transformations---is sufficient to converge to an eigenstate of the Schr{\"o}dinger equation~\cite{mazziotti2004exactness, mazziotti2020}, it provides a well-defined set of actions for RL.  Second, because the residual of the CSE obtained from measurements provides a two-electron description of the current quantum state $S_t(^2D)$ and the update from $S_t(^2D)$ to $S_{t+1}(^2D)$ is Markovian, the gradient-based CQE (Fig. \ref{fig:RL}) can be interpreted as an exploitation-driven algorithm. In the rest of the paper we show that balancing the exploitation with exploration---a key concept of RL algorithms~\cite{Kaelbling.1996, Hasselt.2016}---can significantly accelerate convergence and reduce the overall circuit depth.

\begin{figure*}[t!]
    \centering
    \includegraphics[width=\textwidth]{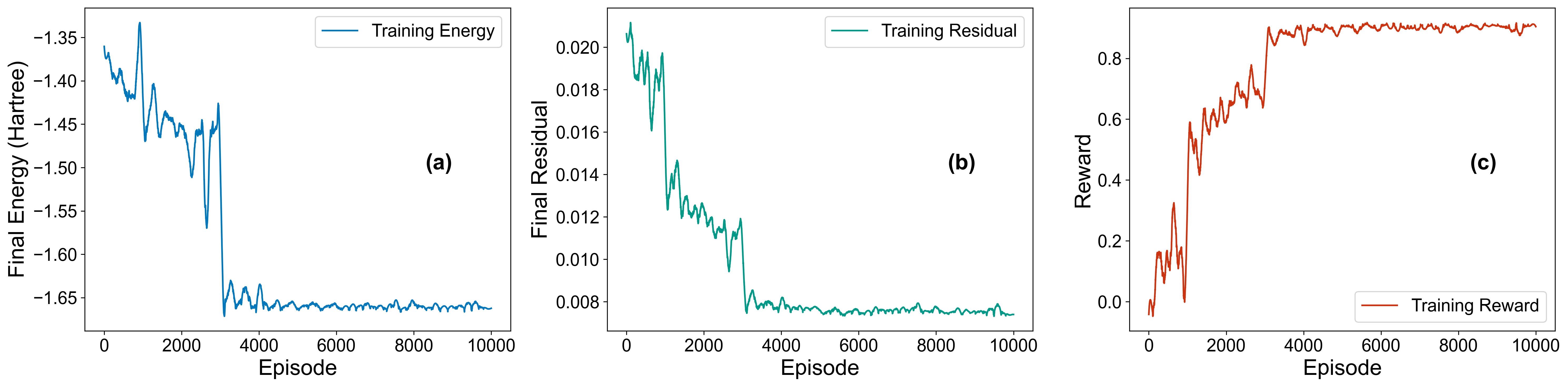}
    \caption{Training metric (a) Ground-state energy estimate, (b) Norm of the CSE residual, and (c) Average reward per episode for H$_4$ with a random geometry. The agent is allowed to take 5~actions.}
    \label{fig:conv}
\end{figure*}

Considering the physics-informed CQE algorithm in Fig. \ref{fig:RL}, the ansatz $T$ after first-order Trotterization can be written as
\begin{equation}
    T(\boldsymbol{\theta}) = \prod_{n}^{N} {\rm e}^{\theta_n \hat{\Gamma}_n}
    \label{eq:ansatz}
\end{equation}
where $\hat{\Gamma}_n$ now registers a single element of the residual tensor in Eq.~(\ref{eq:residual1}). Because the sum in Eq.~(\ref{eq:residual1}) is converted into a product by the Trotterization, reducing ansatz length becomes a minimization of $N$, the total number of exponential transformations. 

In RL our goal is to train an agent described by a state function to select actions from a pool and interact with the environment to maximize the reward. Let us now make the connection between the ansatz optimization problem and RL by defining the essential elements as following:

(1) States: We use the residual of CSE to represent the quantum state as it is a more compact form over wavefunctions while being directly measurable on quantum devices. Furthermore, information about the Hamiltonian is naturally ``absorbed'' into the residual through the commutation relations, which improves its transferability across different molecular geometries. 

(2) Actions: All possible factors in the product form of Eq.~(\ref{eq:ansatz}) generate the pool of actions. Note the operators are reusable in order to parameterize the exact solution~\cite{mazziotti2004exactness,mazziotti2020, evangelista2019}. However, we want to minimize such reuse behavior for circuit depth reduction, which is achieved by adding a penalty term when reused actions are detected.

(3) Environments: Here we use classical computers to simulate the agent-environment interactions as a proof of concept. Additional applications will be achievable by training the agent based on real-time data from quantum computers, which will be investigated in future work.  On a quantum devices, we need to implement only the unitary part of the exponential transformation or employ techniques for effecting non-unitary transformations such as dilation~\cite{Smart.2024, warren2024}

(4) Rewards: It is beneficial in designing the ansatz to minimize not only the energy but also the CSE residual.  To achieve this dual objective, we define a reward function $Q$ that upon maximization will minimize both the energy and the norm of residual $\mathbf{R}$
\begin{equation}
Q = - \left( E + \lambda \|\mathbf{R}\| \right),
\label{eq:reward}
\end{equation}
where $\lambda$ is an adjustable weight factor.

Note that actions consist of both a discrete-indexed reduced density operator $\hat{\Gamma}_{n}$ and a continuous parameter $\theta$. To simplify the learning, we observe that in the original CQE algorithm the parameter $\theta$ is usually optimized in a line-search procedure~\cite{smart2021} or selected as a fixed learning rate~\cite{benavides2024}. This suggests the continuous parameter can be efficiently embedded within a discrete action space. In practice, such an embedding can be implemented by either performing a separate variation of $\theta$ when choosing the action, as in the present work, or predefining a discrete set of allowable $\theta$ values. We employ a classical RL algorithm suitable for handling discrete actions, the dueling double deep Q-network (DQN) framework~\cite{Hasselt.2016, wang2016}.

\section{Results and discussions}

We demonstrate our RL-CQE with the linear molecules H$_3$ and H$_4$ in the Slater-type-orbital with 3 Gaussians (STO-3G) basis set~\cite{hehre1969}.  The Hamiltonians of H$_3$ and H$_4$ are represented with six and eight qubits, respectively, via the Jordan Wigner mapping~\cite{Jordan.1928}.  The spin states are the lowest doublet ($\langle {\hat S}_z \rangle = 1/2$) for H$_3$ and the lowest singlet ($\langle {\hat S}_z \rangle = 0$) for H$_4$.  No additional symmetry constraints are imposed upon the Hamiltonian. The one- and two-electron integrals are obtained from the Quantum Chemistry Package~\cite{rdmchem} for Maple~\cite{maple}.
Our code for the application of RL-CQE to H$_{3}$ is available open source~\cite{RL-CQE2025}.  We show the number of actions to achieve chemical accuracy can be efficiently reduced to single digits with RL.

\subsection{Training Details}

The double DQN framework utilizes an online network for action selection and a target network for stable value estimation in Bellman updates~\cite{Hasselt.2016}. Furthermore, each of these DQNs adopts a dueling architecture, splitting into two streams: one estimating the state value and another estimating the advantage of each action~\cite{wang2016}. These streams are then combined to estimate the Q-values, leading to improved learning efficiency. Structurally, each individual DQN consists of two shared linear layers with Layer Normalization and ReLU activations, branching into two streams each composed of two linear layers and ReLU for value and advantage estimation. We use orthogonal initialization for linear layers and AdamW for optimizing the network.

Training follows a standard deep Q-network approach with prioritized experience replay with a buffer of $5\times10^6$ transitions and proportional prioritization ($\alpha=0.6$) that focuses on learning on experiences with high temporal difference error corrected by importance sampling. A target network, updated every 1000~episodes, and double DQN are implemented to stabilize training and reduce Q-value overestimation. The loss function is a smooth L1 loss between predicted and target Q-values. We use an $\epsilon$-greedy exploration strategy with decaying $\epsilon$ (starting at 1.0, decaying by 0.999 per episode to a minimum of 0.01 unless stated otherwise). Other hyperparameters include a learning rate of $3 \times 10^{-4}$, batch size of 256, a discount factor of 0.99 and a weight factor (Eq. \ref{eq:reward}) of 0.2.

\subsection{Training Performance}

We first present a convergence plot for H$_4$ in Fig.~\ref{fig:conv}. As a proof of concept, our initial guess is a normalized wavefunction generated by assigning each spin-adapted determinant a random-number coefficient. The left (energies), middle (residuals) and right (rewards) panels all demonstrate a similar convergence trend, suggesting effective learning even when the system is initialized with random states. In practice, we also anticipate that employing Hartree-Fock initial guesses would further enhance performance. The residual serves as a more reliable convergence indicator than the energy, especially for systems that are strongly correlated. For example, in the case of H$_4$, we observe that the agent can only converge the residual to less than 0.007 when restricted to five actions, suggesting that there is room for improvement by increasing the number of allowed actions. Furthermore, residuals reflect the effective overlap with the exact solution, which is a more sensitive convergence indicator that can be particularly useful in rugged energy landscapes such as barren plateaus.

\begin{figure}[htbp]
    \centering
    \includegraphics[width=0.9\linewidth]{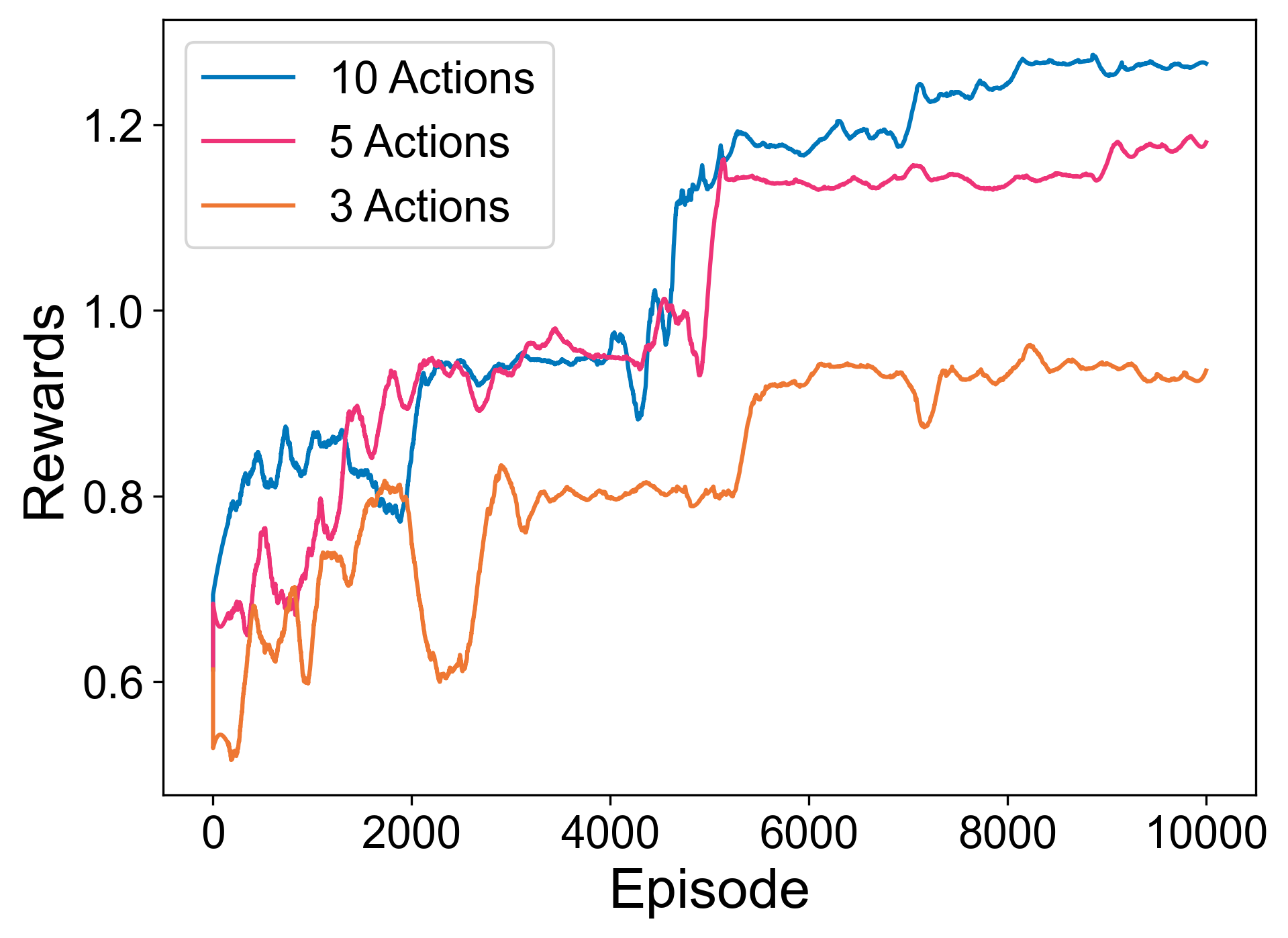}
    \caption{Reward convergence with different numbers of actions for H$_3$ starting from a random wavefunction.}
    \label{fig:step}
\end{figure}

Figure~\ref{fig:step} examines the impact of the ansatz depth (3, 5, and 10 actions) on the convergence, with the 10-action update achieving the highest and most stable rewards. One would naturally expect that increasing the number of actions would lead to a more accurate solution in noiseless environments, as illustrated in this figure. However, we also observe a trade-off between solution accuracy and circuit depth, which can be systematically explored using the RL method presented here. Additionally, when noise effects, such as qubit decoherence, are introduced, RL offers a strategy to explore the best accuracy within specific depth constraints.

\begin{figure}[htbp]
    \centering
    \includegraphics[width=\linewidth]{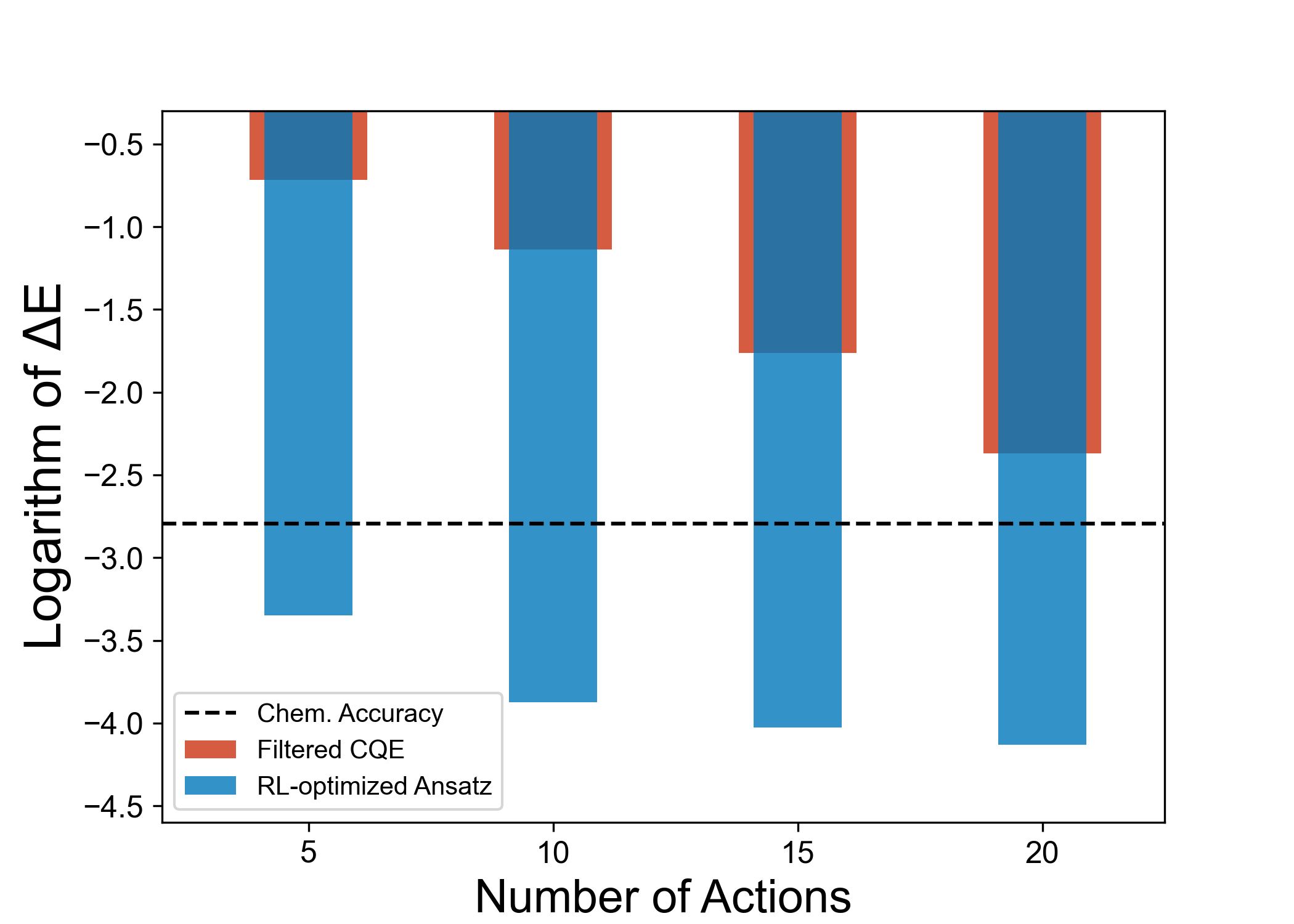}
    \caption{Accuracy and circuit depth for RL optimized ansatz compared to filtered CQE. For filtered CQE at each iteration, the first five actions are applied based on the terms in the transformation that are largest in magnitude. }
    \label{fig:compare}
\end{figure}

We further compare the RL-optimized ansatz with the conventional CQE. A practical technique to reduce the circuit depth of CQE is to filter the residual based on the coefficients after Trotterization~\cite{Smart.2022w8u, wang2024}. This filtering method allows for maintaining a linear increase in depth with iterations while effectively capturing the dominant component of the gradient. Figure~\ref{fig:compare} reports the RL-optimized ansatz compared with filtered CQE where the 5~operators with the largest $^2R^{ij}_{kl}$ coefficients are applied at each iteration. We observe that even with filtering, more than 20 actions are still required for the conventional CQE to reach chemical accuracy, while the RL-optimized ansatz demonstrates a significant and practical improvement in accuracy with the same number of actions. Furthermore, we also observe that the improvement from 10 to 20 actions with RL is trivial, suggesting the molecule can be efficiently simulated with an ansatz of less than 10 exponential transformations. The reason for this significant circuit-depth reduction is that the conventional CQE algorithm often have commuting terms in adjacent iterations that, in principle, can be merged together.  However, in practice, analytically merging these terms is complicated due to the Trotterization error and the specific operator sequence. The proposed RL method, on the other hand, finds a numerically near-optimal solution with given constrains (penalties), which can be used for ansatz optimization on both classical and quantum computers.

\subsection{Model Transferability}

The transferability of the trained model is a critical consideration in the context of designing a universal ansatz. We discuss two scenarios. First, we provide evidence that the model trained on a distribution of Hamiltonians can generate efficient ans\"{a}tze  for a given molecule at different geometries. Second, we discuss the transferability among different devices, i.e. the utilization of classically simulated results for quantum devices and the benefit of training realistic quantum computer data.

\textit{Transferability with different molecular geometries}---The results in Sec.~III~B are trained on the Hamiltonian of a single molecular geometry. For different geometries, retraining or fine-tuning would likely be necessary to ensure optimal performance. To address this challenge, two strategies can be employed. One approach is to leverage transfer learning, where the learned parameters from the original model are used to initialize the training of a new model for the modified Hamiltonian. Another approach is to use a meta-reinforcement learning framework, in which the agent is trained on a distribution of Hamiltonians rather than a single instance. This would allow the model to learn a more generalizable policy that can adapt to different Hamiltonians.

We present some results based on the second methodology. In our experiments, we form a grid of 45 geometries of linear H$_3$ by ranging $1\le r_1\le r_2 \le 5$, where $r_1$ and $r_2$ are the two closest H-H distances in a.u. with the interval being 0.5. The geometry is randomly divided into $k=5$ folds. For each fold, we train a RL model based on resetting each episode with a random Hamiltonian drawn from inside the fold. The trained DQN is then used to evaluate the geometries outside the fold. Table~\ref{tab:crossval} reports the result obtained with the five experiments. It can be seen that the RL agent performs well in predicting the validation set even when a much smaller training set is used. As discussed earlier, one explanation is that using the residual for the input greatly expands the model's transferability.

\begin{table}[h]
    \centering
    \caption{Energy difference between RL optimized ansatz and exact diagonalization. The reported values are averaged over all geometries in the selected training set and the validation set, respectively. The agent is allowed to take 5 actions. }
    \label{tab:crossval}
    \begin{tabular}{ccc}
        \hline
        Sample & Training set (mHa) & Validation set (mHa) \\
        \hline
        1 & 1.23 & 1.61 \\
        2 & 1.21 & 1.87 \\
        3 & 0.92 & 1.21 \\
        4 & 1.31 & 2.25 \\
        5 & 1.76 & 2.44 \\
        \hline
    \end{tabular}
\end{table}


\textit{Transferability between devices}---We are also interested in transferring models trained on one device to another, which includes two major applications. One is to train the model on classical computers to generate an efficient ansatz for practical simulation with quantum computers. In such a scenario, classical computers are used as a resource optimizer for quantum computers. As we have demonstrated in Fig.~\ref{fig:compare}, the ansatz length from RL CQE is well below filtered CQE as well as chemistry-inspired ans\"{a}tze such as unitary coupled cluster ansatz.

Another promising application involves training the RL model using data generated by realistic quantum computers where the RL model can learn the optimal ansatz tailored to the specific noise behavior of the quantum computer. Since many ans\"atze struggle to balance circuit depth, accuracy, and total measurement requirements, the RL protocol provides an efficient framework for identifying the optimal solution under given constraints.

\section{Conclusions and outlook}

In this work we present an RL method to optimize the wavefunction ansatz for many-body molecular simulations on quantum computers. This approach draws inspiration from the CQE algorithm, which iteratively updates the wavefunction through 2-RDM-based measurements on quantum computers. We demonstrate a significant improvement in the convergence rate with RL. The proposed protocol offers an efficient framework for resource optimization and scheduling in quantum simulations on scalable quantum devices. A key difference from some recent work~\cite{nakaji2024, zeng2024, halder2025} in the area of machine-learning-assisted ansatz generation lies in its utilization of the Markovian nature of the CQE, which simplifies the agent's learning process and enables high parallelization.

One potential bottleneck of the current algorithm is that the size of action pool scales polynomially with the number of orbitals. While polynomial scaling is not necessarily problematic, it still can increase the training cost and potentially render efficient training impractical. One future direction is to explore improvements in RL algorithms, such as hierarchical structures or action embeddings, which can reduce the complexity of decision making and improve sample efficiency. Given the rapid progress in the field of RL, further improvements are anticipated to extend beyond the simple yet effective DQN algorithm employed in this study.

While the RL has been proven successful in generating ansatz for different molecular geometries, one further application is to deal with the transferability across different molecules. Models with enhanced generative capabilities such as transformers could be employed for this purpose, enabling the model to move beyond the Markovian process and investigate correlations between operators~\cite{nakaji2024}.  As demonstrated by the theoretical development and calculations, presented here, the RL-CQE methodology provides a significant step towards using AI to accelerate the simulation of ground-state molecular quantum states on existing and emerging quantum devices.


\begin{acknowledgments}
D.A.M. gratefully acknowledges the U.S. National Science Foundation Grant No. CHE-2155082.
\end{acknowledgments}

\bibliography{main}
\end{document}